# Meta-optimization for Fully Automated Radiation Therapy Treatment Planning

Charles Huang, Yusuke Nomura, Yong Yang, and Lei Xing

*Abstract—Objective:* **Radiation therapy treatment planning is a time-consuming process involving iterative adjustments of hyperparameters. To automate the treatment planning process, we propose a meta-optimization framework, called MetaPlanner (MP).** *Methods:* **Our MP algorithm automates planning by performing optimization of treatment planning hyperparameters. The algorithm uses a derivative-free method (i.e. parallel Nelder-Mead simplex search) to search for weight configurations that minimize a meta-scoring function. Meta-scoring is performed by constructing a tier list of the relevant considerations (e.g. dose homogeneity, conformity, spillage, and OAR sparing) to mimic the clinical decision-making process. Additionally, we have made our source code publicly available via github.** *Results:* **The proposed MP method is evaluated on two datasets (21 prostate cases and 6 head and neck cases) collected as part of clinical workflow.  MP is applied to both IMRT and VMAT planning and compared to a baseline of manual VMAT plans. MP in both IMRT and VMAT scenarios has comparable or better performance than manual VMAT planning for all evaluated metrics.** *Conclusion*: **Our proposed MP provides a general framework for fully automated treatment planning that produces high quality treatment plans.** *Significance:* **Our MP method promises to substantially reduce the workload of treatment planners while maintaining or improving plan quality.**

## I. Introduction

RADIOTHERAPY treatment planning is a multi-objective optimization problem that traditionally involves a trial-and-error process for navigating trade-offs [1]. As treatment planning can involve many conflicting objectives (e.g. improve OAR sparing vs. improve PTV coverage), no single plan can optimize performance on all objectives at once. Treatment planning can instead be performed  with the goal of producing Pareto optimal, nondominated solutions—that is we cannot improve one aspect of the plan (e.g., improve sparing in one OAR) without compromising at least one other aspect (e.g., worsen PTV objective) [2]–[4].

At the same time, not all Pareto optimal plans are equally acceptable in the clinic. The main challenge in treatment planning is then to translate our overall clinical goals into weighted objective functions and dose constraints [1]. Following a multicriteria optimization (MCO) approach, we can conceptualize the treatment planning process as navigating the Pareto front in search of the most desirable plans in terms of clinical acceptability. This decision-making process is fundamental to many treatment planning approaches (e.g. manual planning, *a posteriori* MCO, and *a priori* MCO) and is visualized in Figure 1 [4]–[9].

We previously proposed the Pareto optimal projection search (POPS) algorithm to automate treatment planning [4], [6]. Briefly, the POPS algorithm produces treatment plans that are Pareto optimal and clinically acceptable by searching the Pareto front using a treatment plan scoring function. However, the POPS algorithm was only formulated to optimize dose constraint bounds. In this work, we propose a meta-optimization framework that can be adapted to optimize many of the common treatment planning hyperparameters.

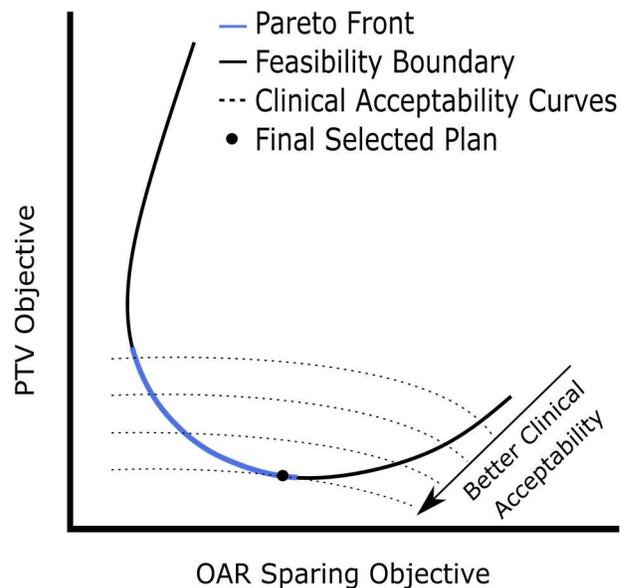

**Figure 1.** Visualization of the decision-making process for human planners in multicriteria optimization

Copyright (c) 2017 IEEE. Personal use of this material is permitted. However, permission to use this material for any other purposes must be obtained from the IEEE by sending an email to pubs-permissions@ieee.org. Manuscript received xxxxxx; revised xxxxxx; accepted xxxxx. Date of publication xxxxxx; date of current version xxxxxx. This work was supported in part by the NIH (1R01 CA176553, 1R01CA256890, R01CA227713, and T32EB009653), as well as a faculty research award from Google Inc.

C. Huang is with the Department of Bioengineering, Stanford University, Stanford, CA 94305.

Y. Nomura, Y. Yang, and L. Xing are with the Department of Radiation Oncology, Stanford University, Stanford, CA 94305 (Ph: (650) 498-7896 E-mail: lei@stanford.edu).

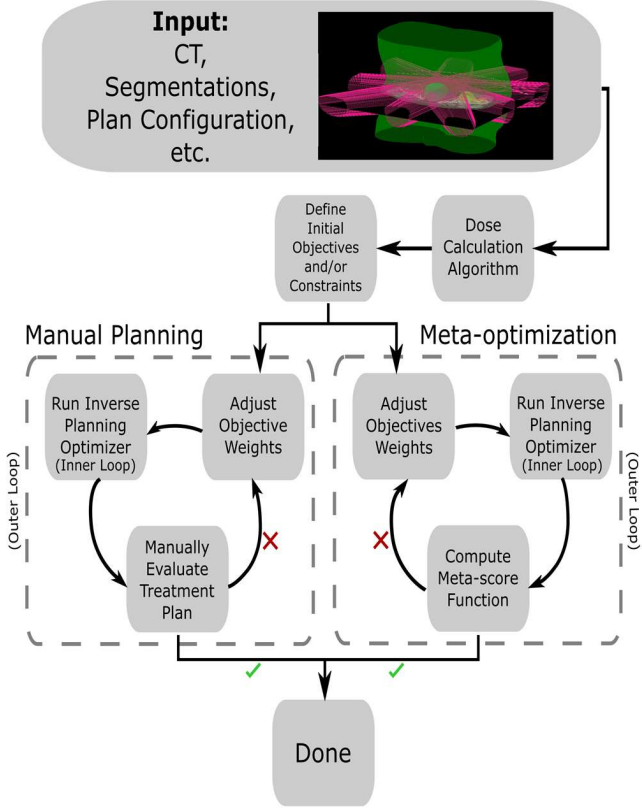

**Figure 2.** Visualization of the meta-optimization workflow in comparison to manual treatment planning.

*A. Problem Formulation and Related Works*

To perform automated treatment planning, we can formulate the problem using an intuitive meta-optimization framework, which we call MetaPlanner (MP). The problem formulation consists of two nested loops of optimization: 1) an inner loop where inverse planning optimization is performed (i.e. fluence map optimization, direct aperture optimization, etc.) and 2) an outer loop where meta-optimization of hyperparameters occurs. This formulation is an automated equivalent to the two loop manual planning approach [10], and the overall workflow for the MP approach is visualized in Figure 2.

Interestingly, we can also demonstrate that the proposed MP approach is a general form of many other previous approaches in MCO [2], [4], [5], [8], [9]. To do so, we first briefly describe the weighted-sum approach, $\epsilon$-constraint method, and the feasibility search, which are popular methods for navigating the Pareto front in MCO methods.

$$\text{Weighted-sum Method}$$
$$\min_{x} \quad w_1 f_1(x) + w_2 f_2(x) + \cdots + w_n f_n(x) \quad (1)$$

The weighted-sum approach begins by assuming that we are given a specific hyperparameter configuration (i.e. objective weights). For this given hyperparameter configuration, we then perform inverse planning optimization until convergence to produce a Pareto optimal treatment plan. Selecting alternative hyperparameter configurations and performing optimization for those configurations results in alternative Pareto optimal plans. When a database of plans with various hyperparameter configurations is created, we are essentially performing *a posteriori* MCO [2], [5]. Traditionally, hyperparameter selection has been performed manually, and we can think of this process as a manual search of Pareto optimal plans.

$$\epsilon\text{-constraint Method}$$
$$\min_{x} \quad f_1(x)$$
$$s.t. \quad 0 \leq f_2(x) \leq c_2$$
$$\vdots$$
$$0 \leq f_n(x) \leq c_n$$
$$x \in X \quad (2)$$

Unlike the weighted-sum approach, where we have multiple objectives and weights, the $\epsilon$-constraint method considers a single objective function and converts other criteria into constraints. Many *a priori* MCO approaches build on the $\epsilon$-constraint method. For instance, in *a priori* MCO approaches using the lexicographic method, we sequentially perform inverse planning optimization with the $\epsilon$-constraint method in the order of priority given by a list preferences or rules, iteratively converting each objective into an additional constraint [8], [9].

A third method, the feasibility search, instead assumes that we are given a specific hyperparameter configuration in the form of only dose constraint bounds. Performing inverse planning optimization with a specific hyperparameter configuration (i.e. set of dose constraints) allows for testing of constraint feasibility. We can then determine the feasibility boundary (which contains the Pareto front) by iteratively tightening our constraint bounds until we reach the boundary between feasible and infeasible treatment plans. MCO methods that utilize the feasibility search include the Pareto Optimal Projection Search (POPS) method and the Noncoplanar-POPS method (NC-POPS) [4], [6].

In essence, these three popular methods for MCO are alternative ways of navigating the Pareto front to find treatment plans that are clinically acceptable. MCO methods differ primarily in how navigation of the Pareto front is performed. Many traditional MCO approaches like *a posteriori* methods are only semi-automated, as final hyperparameter selection is still performed manually. Other MCO methods, such as POPS, are fully automated but only work with dose constraints. In this work, we extend on previous MCO approaches and propose a general framework for fully automated MCO of treatment plans, called MetaPlanner (MP). Instead of navigating the Pareto front manually or using predefined rules, meta-optimization utilizes an additional layer of optimization to navigate the Pareto front. This additional layer of meta-optimization is more efficient in searching the Pareto front than heuristic navigation, more automated than traditional *a posteriori* MCO methods, and more general than methods like POPS, which rely only on dose constraints. The implementation of the proposed MP approach is described below.

## II. METHODS

### A. MetaPlanner Implementation

The proposed MP approach to automated planning performs meta-optimization on plan hyperparameters using two nested optimization loops. The inner optimization loop consists of any traditional inverse optimization approach for treatment planning (i.e. FMO, DAO, etc.). In the outer optimization loop, MP uses the parallel Nelder-mead simplex search algorithm [11], [12] to optimize plan hyperparameters. An example workflow for the MP approach is visualized in Figure 3 for two popular treatment modalities in external beam therapy (i.e. IMRT and VMAT). In this current implementation, only fluence map optimization is used for the inner loop in order to reduce computation time. The rest of the workflow follows the default pipeline provided by the MatRad treatment planning software package, which sequentially performs FMO, leaf sequencing, and DAO for VMAT plans and FMO followed by leaf sequencing for IMRT plans [13], [14]. In the proposed MP approach, we formulate the meta-optimization problem following Equation 3.

$$\text{MetaPlanner (MP)}$$
$$\min_{w} \quad f_{meta}(w)$$
$$s.t. \quad w \geq 0$$
$$\mathbf{1}^T w \leq 1 \quad (3)$$

Here, $w$ refers to the objective function weights for inverse planning optimization and $f_{meta}$ refers to the meta-scoring function used to evaluate the clinical acceptability of each treatment plan.

### B. Meta-scoring of Treatment Plans

Attempting to mimic clinical decision-making, we adopt a tier list for ranking planner preferences. This tier list method makes two main assumptions. First, we assume that tiers follow an ascending order where lower tiers have greater importance (i.e. $\tau_0 \gg \tau_1 \gg \tau_2 \ldots \gg \tau_m$). Second, we assume that considerations within each tier have similar or equal importance.

Meta-scoring of plans also utilizes many dose statistics and indices that are routinely used in clinical decision-making: dose homogeneity [15], [16], dose conformity [17], [18], dose spillage [19], and mean OAR dose. Desired ranges (i.e. notations containing $\{(-), (+)\}$) are adapted from standard protocol (i.e. RTOG 0126 and NRG HN005).

- We first define the homogeneity index (HI) as the following: $HI = 100 \times \frac{D_5 - D_{95}}{D_p}$. Here, $D_5$ refers to the dose received by 5% of the target volume, $D_{95}$ refers to the dose received by 95% of the target volume, $D_p$ refers to the prescription dose, and $\{HI_{(-)}, HI_{(+)}\}$ refers to the desired range for homogeneity.

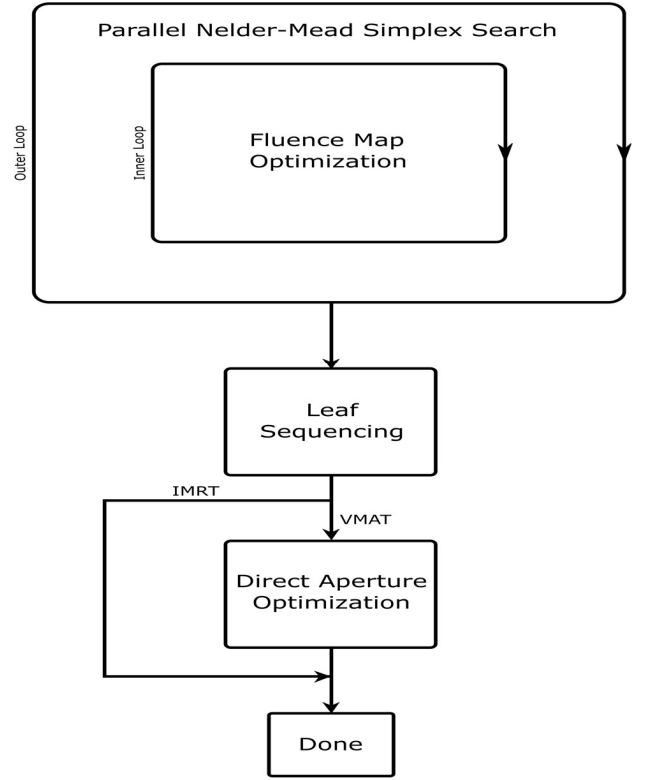

**Figure 3.** Visualization of the MP approach applied to IMRT and VMAT planning.

- We use the following definition for dose conformity: $CI = \frac{(TV_{95\ p})^2}{TV \times V_{95\ p}}$. Here $TV_{95D_p}$ refers to the target volume covered by 95% of the prescription dose, TV refers to the target volume, $V_{95D_p}$ refers to the volume covered by 95% of the prescription dose, and $\{CI_{(-)}, CI_{(+)}\}$ refers to the desired range for conformity.
- We use the following definition for 90% dose spillage: $R90 = \frac{V_{90D_p}}{TV}$. Here TV refers to the target volume, $V_{90D_p}$ refers to the volume covered by 90% of the prescription dose, and $\{R90_{(-)}, R90_{(+)}\}$ refers to the desired range for R90.
- We use the following definition for 50% dose spillage: $R50 = \frac{V_{50\ p}}{TV}$. Here TV refers to the target volume, $V_{50\ p}$ refers to the volume covered by 50% of the prescription dose, and $\{R50_{(-)}, R50_{(+)}\}$ refers to the desired range for R50.
- For OAR sparing, we compute the mean dose as follows: $\bar{d}_s = \frac{1}{n_s}\sum_{i \in s}^{n_s} d_i$. Here, $d_i$ refers to the dose at voxel $i$ and $n_s$ refers to the number of voxels in structure $s$. $\{\bar{d}_{s,(-)}, \bar{d}_{s,(+)}\}$ refers to the desired range for the mean dose.

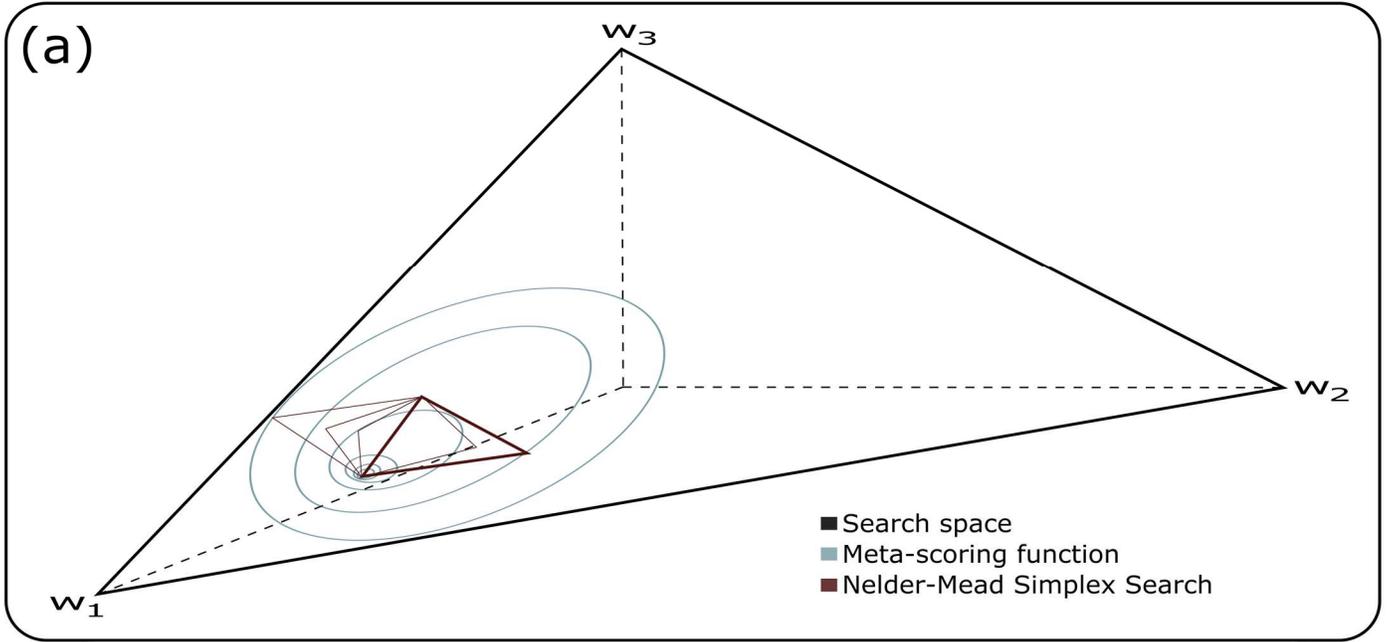

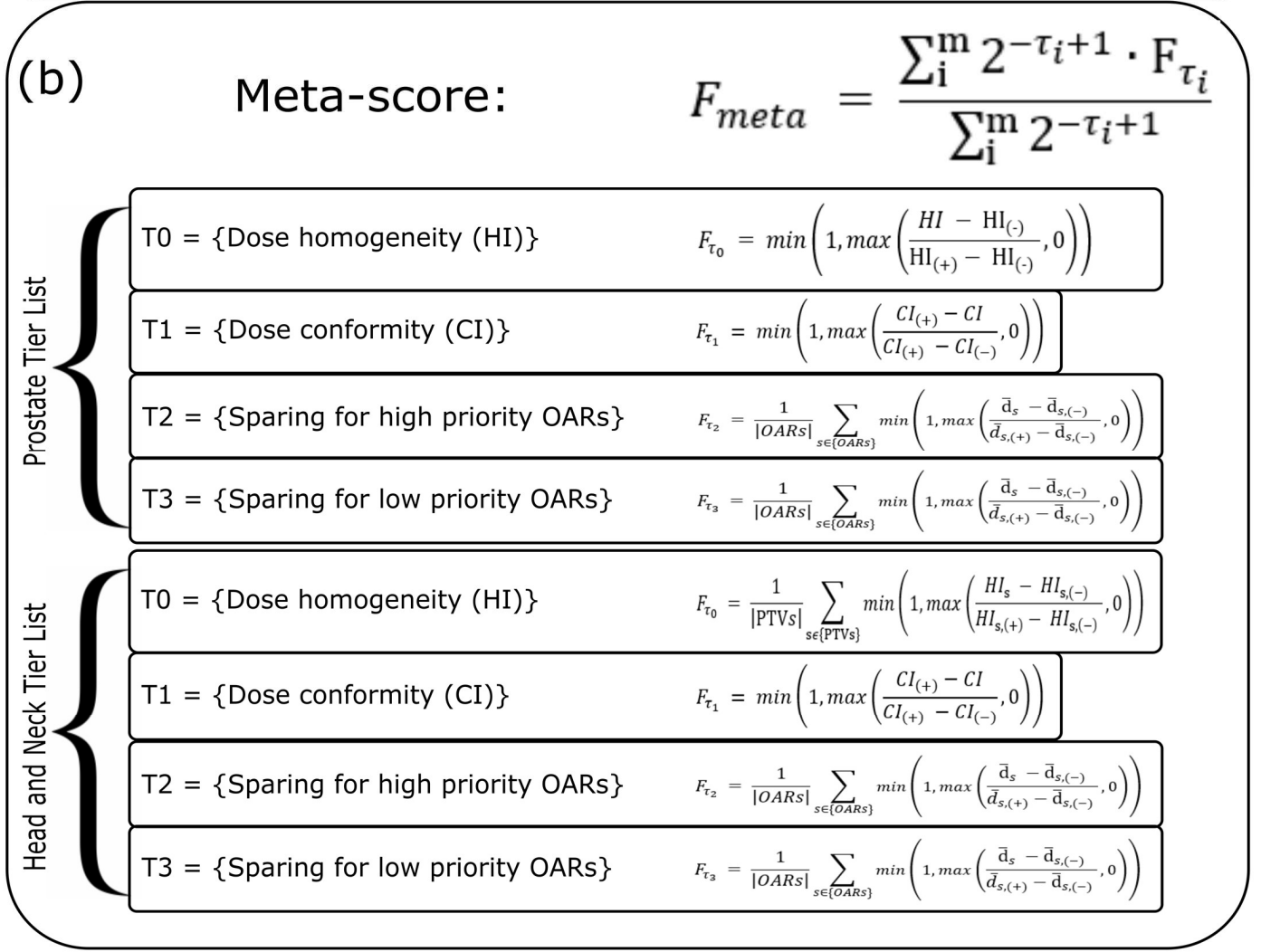

**Figure 4.** (a) Visualizes the parallel Nelder-mead simplex search which is used to meta-optimize objective function weights. (b) Defines the meta-scoring function $F_{meta}$, along with the tier list used incorporate planner preferences. A separate tier list is used for prostate and head and neck cases.

For meta-scoring of treatment plans, we can construct each tier list term $F_{\tau_i}$ by averaging all considerations in each tier. As an example, when computing $F_{\tau_i}$ for $\tau_i = 0$ in head and neck cases, we first normalize the three homogeneity terms (HI$_{PTV\,70}$, HI$_{PTV\,56}$, and HI$_{PTV\,52}$) to the range [0,1] and then compute their average. The meta-scoring function $F_{meta}$ is then computed as the weighted average of each tier list term $F_{\tau_i}$. As $\tau_i$ increases (decreasing importance), the weight applied to $F_{\tau_i}$ decreases. Figure 4b provides definitions for each tier list term $F_{\tau_i}$ and the overall meta-scoring function $F_{meta}$.

*C. MatRad Implementation*

The proposed MP approach utilizes the open-source MatRad software package (dev_VMAT branch) [13], [14], [20]. Dose calculation (i.e. computation of the dose-influence matrix) was performed using a singular value decomposed pencil beam algorithm [21]. Due to memory limitations with the current MatRad version, all plans used a pencil beam size of $5 \times 5\ mm^2$ and a voxel size of $3.5 \times 3.5 \times 3.5\ mm^3$. For MP IMRT plans, we selected a configuration of 7 and 9 equally spaced beams for prostate and head and neck cases, respectively. MP VMAT plans were generated with dual arcs (720°) following the MatRad implementation of the SmartArc planning algorithm [14], [20], which performs FMO, aperture sequencing, and DAO. Inverse planning optimization (i.e. FMO and DAO) uses the interior-point optimization (IPOPT) package to solve the problem formulated in Equation 4. Here, $w$ refers to the objective function weights, $d_i$ refers to the dose at voxel $i$, $\hat{d}$ refers to the structure-specific reference dose, $\Theta(\cdot)$ refers to the Heaviside function, $N_s$ refers to the number of voxels in structure $s$, $\{\cdot\}$ refers to the set of OARs or PTVs, and $\boldsymbol{D}$ refers to the dose-influence matrix. For FMO, $x$ refers to the pencil beam weights. For DAO, the optimization problem description is provided in the original MatRad papers [13], [20]. Table 1 summarizes the meta-scoring tier list and other relevant parameters. While we cannot release our datasets at this time, we have uploaded our source code and made it publicly available via github (https://github.com/chh105/MetaPlanner). Other parameters not listed here can be found in the source code.

### III. RESULTS

*A. Experimental Setup and Evaluation*

To evaluate the proficiency of our proposed MP approach, we perform a retrospective comparison on a dataset of 21 prostate cases and a dataset of 6 head and neck cases (Stanford IRB protocol #41335). In particular, we compare MP generated IMRT and VMAT treatment plans to manual VMAT treatment

Inverse Planning Formulation

$$\min_{x}\ w_{PTV} \sum_{s \in \{PTVs\}} \frac{1}{N_s} \sum_{i \in s}(d_i - \hat{d})^2 + \sum_{s \in \{OARs\}} \frac{w_s}{N_s} \sum_{i \in s} \Theta(d_i - \hat{d})(d_i - \hat{d})$$

$$s.t. \quad x \geq 0$$
$$\vec{d} = \boldsymbol{D}\vec{x} \qquad (4)$$

**Table 1.** Lists the inverse planning problem formulation, meta-scoring tier list, and other relevant parameters used. Here, we are given a set of hyperparameters ($\vec{w}$), which are updated by the outer optimization loop of the MP approach.

|  | Tier 0 | Tier 1 | Tier 2 | Tier 3 |  | Overlap Priority | Reference Dose ($\hat{d}$) |
|---|---|---|---|---|---|---|---|
| Prostate Cases | HI | CI | Sparing (Rectum, Bladder) | Sparing (FH R, FH L, Body) | PTV | 1 | 74 |
|  |  |  |  |  | Rectum | 2 | 0 |
|  |  |  |  |  | Bladder | 2 | 0 |
|  |  |  |  |  | FH R | 2 | 30 |
|  |  |  |  |  | FH L | 2 | 30 |
|  |  |  |  |  | Body | 2 | 30 |
| Head and Neck Cases | HI (PTV 70, PTV 56, PTV 52) | CI (PTV 70) | Sparing (Spinal Cord, Brainstem, Parotid R, Parotid L, Oral Cavity, Larynx,) | Sparing (Body) | PTV 70 | 1 | 72 |
|  |  |  |  |  | PTV 56 | 2 | 58 |
|  |  |  |  |  | PTV 52 | 3 | 54 |
|  |  |  |  |  | Cord | 4 | 30 |
|  |  |  |  |  | Brainstem | 4 | 30 |
|  |  |  |  |  | Parotid R | 4 | 0 |
|  |  |  |  |  | Parotid L | 4 | 0 |
|  |  |  |  |  | Oral Cavity | 4 | 0 |
|  |  |  |  |  | Larynx | 4 | 0 |
|  |  |  |  |  | Body | 4 | 30 |

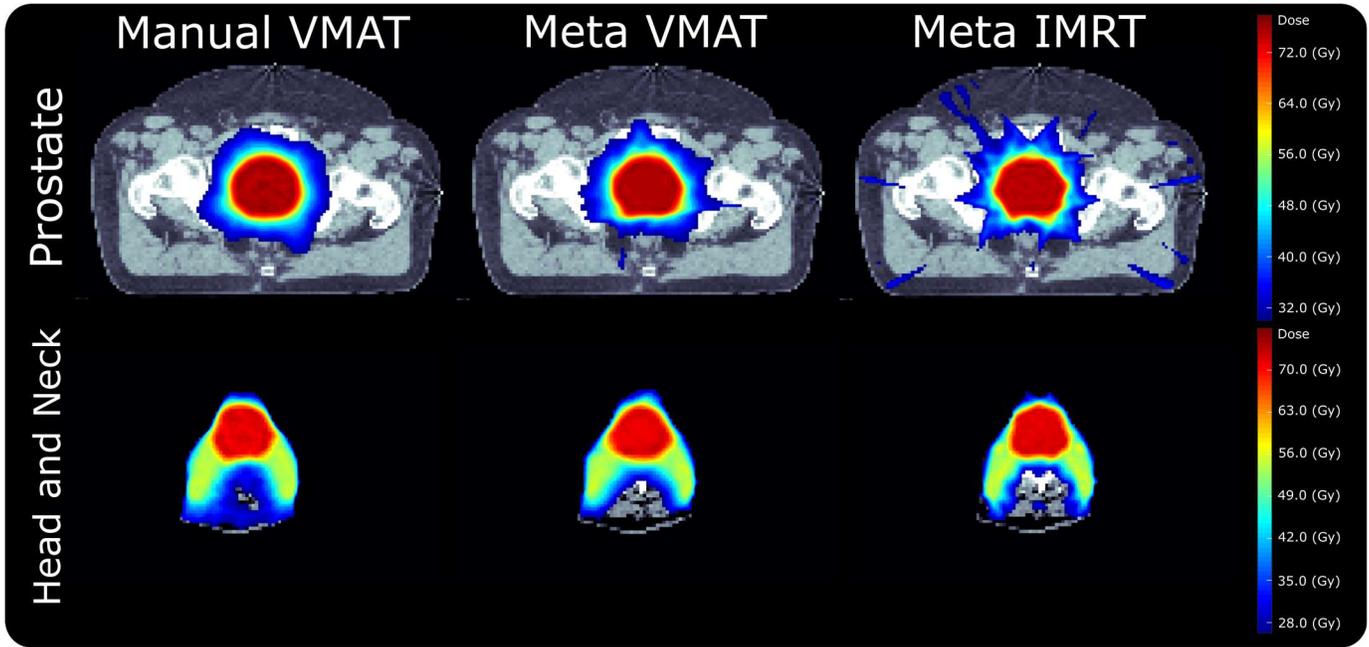

**Figure 5.** Visual comparison of dose distributions for a representative prostate case and a representative head and neck case.

plans originally created as part of clinical workflow. These manual VMAT plans were planned by experienced planners using the Eclipse treatment planning software from Varian Medical Systems, where the MP approach has not yet been implemented. Additionally, these manual plans use a configuration of two full coplanar arcs.

*B. Qualitative Comparison*

We first perform a qualitative comparison between plans produced by our MP approach and the baseline manual plans. Figure 5 provides a visualization of the differences in dose distributions between the various approaches for a representative prostate case and a representative head and neck case. Visually, all three methods produce highly conformal and homogeneous plans. For the prostate case, sparing for OARs like the femoral heads and body is also significantly better for the MP plans than the manual plan. For the head and neck case, dose spillage is improved for the MP IMRT and MP VMAT methods.

A visualization of the mean dose-volume histograms, with corresponding standard deviations shown as error bands, is provided in Figure 6. For prostate cases, the DVHs for the rectum and bladder are comparable for the three methods. However, the DVHs for the femoral heads and body demonstrate better sparing for the MP IMRT and MP VMAT methods, as compared to the manual planning baseline. For head and neck cases, the DVHs for OARs appear visually comparable for the MP VMAT and manual VMAT methods, and the DVHs for the MP IMRT method demonstrate better sparing for most OARs (i.e. cord, brainstem, oral cavity, etc.). The DVHs of manual plans for PTV 52 and PTV 56 additionally appear to have greater variability in dose homogeneity, as compared to MP VMAT and MP IMRT plans.

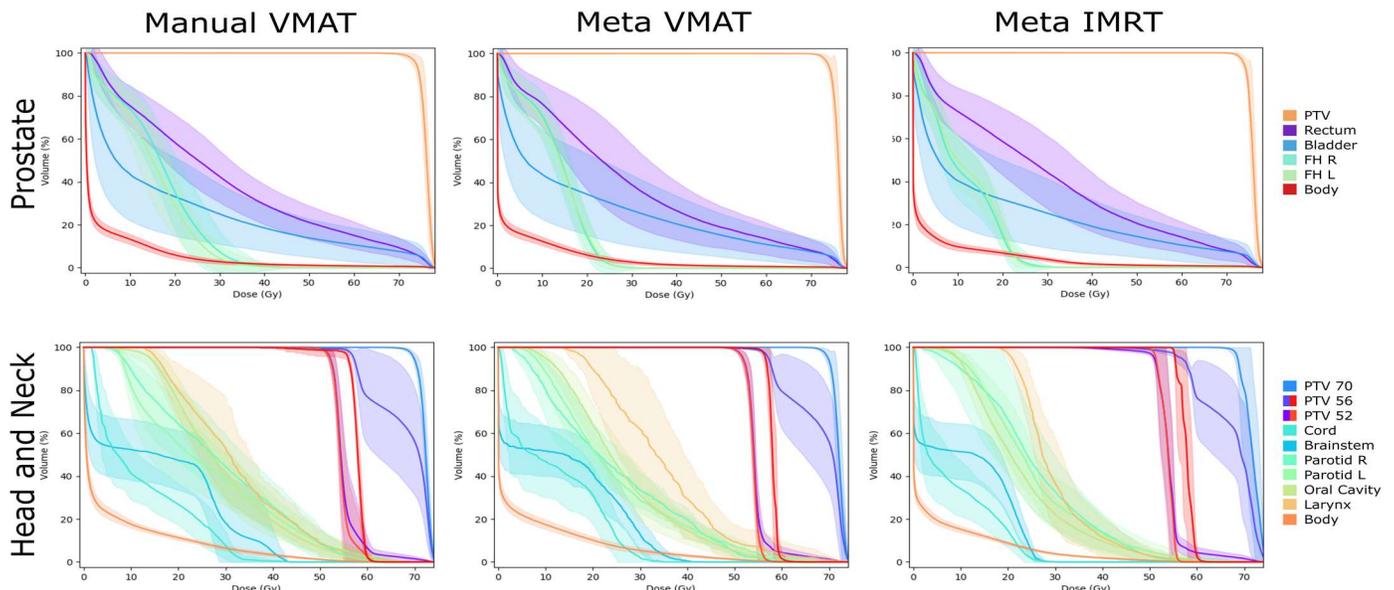

**Figure 6.** Visual comparison of the dose-volume histograms for both the prostate and head and neck datasets. Here, the means are shown as solid lines and the standard deviations are shown as error bands.

*C. Quantitative Comparison*

Tables 2 and 3 provide further quantitative comparison between the baseline manual VMAT plans, MP VMAT plans, and MP IMRT plans. Differences in dose conformity, dose homogeneity, dose spillage (i.e. R90 and R50), and OAR sparing (mean dose) are quantified using the Wilcoxon signed-rank test ($p < 0.05$).

*1) Prostate Cases*

Dose conformity values are summarized in Table 2. Here, CI values approach 1 for an ideal case. IMRT and VMAT plans generated using the proposed MP approach were significantly more conformal than manual VMAT plans, with p-values of 0.00001 and 0.00226, respectively. Moreover, MP IMRT plans were significantly more conformal than MP VMAT plans (p-value of 0.00001). Dose homogeneity values are also evaluated in Table 2, where HI approaches 0 for an ideal case. Here, all three methods produced plans of comparable dose homogeneity. Dose spillage is evaluated using R90 and R50. For R90, both MP IMRT and MP VMAT performed significantly better than manual VMAT planning (p-values of 0.00001 and 0.00306, respectively). MP IMRT also had significantly better R90 performance than MP VMAT (p-value of 0.00012). For R50, MP VMAT performed significantly better than manual VMAT planning and MP IMRT (p-values of 0.00155 and 0.00019, respectively). MP IMRT and manual VMAT had comparable R50 performance.

Similarly, OAR sparing was evaluated for the rectum, bladder, femoral heads, and body. The rectum received comparable sparing for MP IMRT and manual VMAT but significantly better sparing for MP VMAT (p-value of 0.00636). The bladder received comparable sparing for MP VMAT and manual VMAT but significantly better sparing for MP IMRT (p-value of 0.00048). Both femoral heads had significantly better sparing when using MP IMRT and MP VMAT, as compared to manual VMAT. Finally, for body sparing, MP VMAT and MP IMRT had significantly better body sparing than manual VMAT (p-values of 0.00008 and 0.00009).

**Table 2.** Quantitative comparison is performed for prostate cases, where we assess dose homogeneity, conformity, spillage, and OAR sparing. Methods are compared using the Wilcoxon signed-rank test ($p < 0.05$) with the best values bolded for readability.

| | Conformity Index (CI) | Homogeneity Index (HI) | R90 | R50 | OAR | Mean Dose ($\mu$) | D(2%) (Gy) | D(20%) (Gy) | D(40%) (Gy) |
|---|---|---|---|---|---|---|---|---|---|
| Manual VMAT | 0.86 (0.03) | 4.63 (0.98) | 1.29 (0.05) | 3.44 (0.20) | Rectum | 30.0 (5.5) | 75.3 (2.6) | 51.0 (10.4) | 31.1 (7.5) |
| | | | | | Bladder | 19.1 (9.5) | 73.0 (6.7) | 35.4 (21.0) | 14.9 (11.9) |
| | | | | | FH R | 16.3 (3.4) | 31.4 (5.3) | 24.0 (4.8) | 19.5 (4.3) |
| | | | | | FH L | 14.9 (3.1) | 30.6 (5.2) | 22.1 (4.6) | 17.1 (3.4) |
| | | | | | Body | 3.9 (0.8) | 35.1 (6.7) | 3.2 (1.6) | 0.5 (0.2) |
| MP VMAT | 0.90 (0.01) | 4.14 (0.62) | 1.24 (0.03) | **3.26 (0.17)** | Rectum | **28.6 (8.2)** | 74.1 (6.8) | 46.8 (13.4) | 30.1 (10.1) |
| | | | | | Bladder | 19.6 (10.2) | 73.6 (6.7) | 36.9 (20.3) | 16.0 (12.8) |
| | | | | | FH R | 13.1 (2.0) | 24.2 (2.8) | 19.3 (1.7) | 16.9 (1.9) |
| | | | | | FH L | 13.0 (1.6) | 24.8 (2.2) | 19.5 (1.9) | 16.2 (2.0) |
| | | | | | Body | 3.6 (0.7) | 35.2 (5.6) | 4.1 (4.0) | 0.5 (1.5) |
| MP IMRT | **0.93 (0.01)** | 4.11 (0.69) | **1.19 (0.02)** | 3.58 (0.28) | Rectum | 29.5 (7.7) | 74.0 (6.9) | 46.2 (13.8) | 28.8 (10.0) |
| | | | | | Bladder | **18.5 (9.9)** | 73.8 (5.7) | 37.1 (20.0) | 15.6 (13.5) |
| | | | | | FH R | **10.7 (2.2)** | 23.9 (2.3) | 18.8 (1.9) | 15.6 (2.1) |
| | | | | | FH L | **11.2 (1.9)** | 24.4 (2.0) | 19.1 (1.7) | 15.7 (1.6) |
| | | | | | Body | **3.4 (0.7)** | 34.6 (5.6) | 2.3 (1.4) | 0.1 (0.0) |

| | | Conformity Index (CI) | Homogeneity Index (HI) | R90 | R50 | Rectum ($\mu$) | Bladder ($\mu$) | FH R ($\mu$) | FH L ($\mu$) | Body ($\mu$) |
|---|---|---|---|---|---|---|---|---|---|---|
| MP VMAT vs. Manual VMAT | Wilcoxon Signed-rank Test (p-value) | **0.00226** | 0.22883 | **0.00306** | **0.00155** | 0.23047 | 0.39446 | **0.00028** | **0.00961** | **0.00008** |
| MP IMRT vs. Manual VMAT | Wilcoxon Signed-rank Test (p-value) | **0.00001** | 0.13037 | **0.00001** | 0.13962 | 0.52021 | 0.18084 | **0.00006** | **0.00016** | **0.00009** |
| MP VMAT vs. MP IMRT | Wilcoxon Signed-rank Test (p-value) | **0.00001** | 0.10247 | **0.00012** | **0.00019** | **0.00636** | **0.00048** | **0.00006** | **0.00019** | **0.00187** |

Overall, the best performance for dose conformity, dose spillage (R90 and R50), rectum sparing, bladder sparing, sparing of the femoral heads, and body sparing was achieved using either MP IMRT or MP VMAT. Dose homogeneity was comparable for the three methods.

*2) Head and Neck Cases*

Quantitative results for head and neck cases are summarized in Table 3. We first compare MP VMAT to baseline manual VMAT planning. Performance for dose conformity, dose homogeneity, and OAR sparing of all organs except the body is comparable for both methods. However, MP VMAT provides a significant reduction in low dose spillage as compared to manual VMAT, resulting in a lower mean dose to the body.

We similarly compare MP IMRT to baseline manual VMAT planning. MP IMRT achieves significantly better performance for dose conformity, dose homogeneity to PTV 56 and PTV 70, and OAR sparing for all organs except the brainstem and left parotid. For dose homogeneity to PTV 52 and OAR sparing of the remaining organs (i.e. brainstem and left parotid), performance was comparable between MP IMRT and manual VMAT.

Finally, we also compare MP VMAT and MP IMRT. For dose conformity, dose homogeneity for PTV 52 and PTV 70, and OAR sparing of all organs except the parotids and the body, MP IMRT significantly outperforms MP VMAT. Dose homogeneity for PTV 56 and OAR sparing of the parotids and body were comparable for the two methods.

## IV. DISCUSSION

This study proposed the MP approach for automated treatment planning. MP is a novel two loop meta-optimization approach that optimizes treatment plan hyperparameters (e.g. objective weights). Here, we apply our proposed MP approach to automated planning for IMRT and VMAT, and we compare its performance to a baseline method of manual VMAT planning. The proposed approach produces treatment plans that have comparable or better performance to manual plans in all evaluated metrics (dose conformity, homogeneity, spillage, OAR sparing, etc.).

**Table 3.** Quantitative comparison is performed for head and neck cases, where we assess dose homogeneity, conformity, and OAR sparing. Methods are compared using the Wilcoxon signed-rank test ($p < 0.05$) with the best values bolded for readability.

| | CI (PTV 70) | HI (PTV 52) | HI (PTV 56) | HI (PTV 70) | OAR | Mean Dose ($\mu$) | D(2%) (Gy) | D(20%) (Gy) | D(40%) (Gy) |
|---|---|---|---|---|---|---|---|---|---|
| Manual VMAT | 0.83 (0.02) | 7.40 (3.27) | 3.42 (1.15) | 5.22 (0.57) | Cord | 15.7 (6.5) | 33.8 (4.6) | 29.7 (5.2) | 22.1 (11.4) |
| | | | | | Brainstem | 12.0 (2.7) | 32.3 (4.3) | 22.2 (6.9) | 12.0 (5.2) |
| | | | | | Parotid R | 28.6 (6.1) | 55.8 (6.9) | 42.3 (6.0) | 31.3 (7.0) |
| | | | | | Parotid L | 26.2 (5.7) | 58.8 (6.4) | 40.7 (5.0) | 27.0 (8.3) |
| | | | | | Oral Cavity | 29.2 (3.2) | 55.9 (3.7) | 39.0 (4.0) | 30.0 (3.9) |
| | | | | | Larynx | 32.0 (4.6) | 57.2 (5.8) | 41.9 (5.8) | 32.5 (4.9) |
| | | | | | Body | 5.7 (0.7) | 45.5 (2.5) | 7.4 (2.0) | 1.1 (0.3) |
| MP VMAT | 0.82 (0.05) | 6.03 (1.46) | 2.66 (2.43) | 4.34 (0.72) | Cord | 14.1 (4.6) | 35.8 (4.2) | 27.9 (4.0) | 18.5 (9.3) |
| | | | | | Brainstem | 11.1 (4.0) | 28.4 (2.5) | 20.6 (7.3) | 13.1 (6.6) |
| | | | | | Parotid R | 22.9 (4.3) | 52.4 (4.8) | 34.8 (5.2) | 23.3 (5.7) |
| | | | | | Parotid L | 24.1 (2.4) | 56.9 (6.5) | 34.5 (4.4) | 23.7 (3.9) |
| | | | | | Oral Cavity | 27.1 (2.5) | 55.7 (3.9) | 37.5 (4.1) | 27.7 (2.7) |
| | | | | | Larynx | 34.3 (5.1) | 54.3 (6.7) | 41.7 (5.3) | 35.2 (5.4) |
| | | | | | Body | 5.0 (0.5) | 44.3 (2.9) | 6.5 (1.9) | 0.2 (0.1) |
| MP IMRT | **0.85 (0.03)** | **5.02 (1.08)** | **2.11 (1.66)** | **3.50 (0.46)** | Cord | **12.0 (2.8)** | 31.3 (1.9) | 23.5 (1.7) | 16.1 (7.2) |
| | | | | | Brainstem | **10.0 (3.7)** | 27.3 (3.3) | 20.3 (7.5) | 10.0 (7.2) |
| | | | | | Parotid R | **20.7 (3.3)** | 48.7 (3.6) | 32.4 (5.5) | 21.3 (4.2) |
| | | | | | Parotid L | **23.0 (1.1)** | 55.3 (5.5) | 34.6 (2.9) | 23.1 (2.9) |
| | | | | | Oral Cavity | **24.0 (2.7)** | 53.7 (4.1) | 33.9 (3.9) | 24.5 (3.7) |
| | | | | | Larynx | **26.8 (3.6)** | 49.9 (6.2) | 34.2 (4.9) | 26.7 (4.4) |
| | | | | | Body | **4.7 (0.5)** | 42.1 (2.3) | 5.6 (2.1) | 0.2 (0.1) |
| | | CI (PTV 70) | HI (PTV 52) | HI (PTV 56) | HI (PTV 70) | Cord ($\mu$) | Brainstem ($\mu$) | Parotid R ($\mu$) | Parotid L ($\mu$) | Oral Cavity ($\mu$) | Larynx ($\mu$) | Body ($\mu$) |
| MP VMAT vs. Manual VMAT | Wilcoxon Signed-rank Test (p-value) | 0.46307 | 0.60017 | 0.34544 | 0.14111 | 0.17295 | 0.34545 | 0.07474 | 0.46307 | 0.11585 | 0.46307 | **0.02771** |
| MP IMRT vs. Manual VMAT | Wilcoxon Signed-rank Test (p-value) | **0.04639** | 0.11585 | **0.04639** | **0.02771** | **0.02771** | 0.17295 | **0.0464** | 0.34544 | **0.02771** | **0.04639** | **0.02771** |
| MP VMAT vs. MP IMRT | Wilcoxon Signed-rank Test (p-value) | **0.02771** | **0.02685** | 0.17295 | **0.02771** | **0.0464** | **0.0464** | 0.07474 | 0.17295 | **0.0464** | **0.02771** | 0.17295 |

As discussed in previous sections, many MCO approaches have been proposed in the past for treatment planning, with varying degrees of automation. *A posteriori* methods, for instance, are semi-automated methods that produce a database of Pareto optimal plans and leave the final plan selection to a human planner [2], [5]. Other works, like the POPS algorithm, are fully automated but are formulated to only handle dose constraints [4], [6]. The proposed method, MetaPlanner, is a general framework for fully automated MCO, and it can be adapted to optimize any of the common treatment planning hyperparameters (i.e. dose objective weights, dose constraint bounds, etc.).

One important component of the MP approach is the search algorithm. The current implementation of the MP approach uses the parallel Nelder-mead simplex search routine, which performs a derivative-free search of the objective function weights using multiple workers. Like other derivative-free methods, the Nelder-mead simplex search often performs well for problems with expensive function evaluations and a relatively small number of optimization variables [22], making it a good candidate for automated treatment planning.

The proposed MP approach requires no active planning and has relatively fast computation times, performing meta-optimization in around an hour ($52.7 \pm 13.6$ min). DAO, which was used in VMAT planning, required an additional hour ($77.1 \pm 13.9$ min). MP results were tested on a consumer desktop, which uses a Ryzen 2700x (3.7 GHz). Most of the computation in the proposed approach is spent performing inverse planning optimization, so improving inverse planning optimization speed could greatly reduce overall planning times. Recently, Macfarlane et al. [14] proposed the fast inverse direct aperture optimization algorithm (FIDAO) and reported between a 7 to 32-fold speedup over the conventional DAO algorithm used by MatRad. Methods like FIDAO reduce the computation times of treatment planning in MatRad, allowing for similar computation times to those of clinical software implementations [14], [23].

In this study, we performed a retrospective evaluation on a dataset of prostate cases and proposed the MP approach for fully automated treatment planning. Moving forward, we hope to extend the MP approach to other body sites (i.e. abdomen, lungs, etc.) through modifications to the meta-scoring function. Separately, noncoplanar treatment planning has the potential to improve dosimetric quality, as shown by previous studies in literature [6], [24], [25]. We hope to adapt the MP approach to noncoplanar planning in future studies as well.

## V. Conclusion

External beam radiation therapy is used for treatment of a substantial number of cancer patients [26]. In this work, we proposed the MetaPlanner (MP) framework for automated treatment planning, and applied the proposed method to both prostate and head and neck cases. Our experimental results demonstrate that the proposed MP method performs comparable to or better than manual planning for all evaluated metrics (i.e. dose conformity, dose homogeneity, OAR sparing, etc.). We anticipate that the proposed method will improve treatment planning workflow and elevate plan quality. Additionally, our source code has been made publicly available on github.

## VI. Acknowledgement

The authors would like to thank and acknowledge the MatRad development team for their help and advice.